\documentclass{aastex631}

\begin{document}

\title{Image Pre-Processing Framework for Time-Domain Astronomy in the Artificial Intelligence Era}

\correspondingauthor{Peng Jia}
\email{robinmartin20@gmail.com}

\author{Liang Cao}
\affiliation{College of Physics and Optoelectronics, Taiyuan University of Technology, Taiyuan, 030024, China}

\author[0000-0001-6623-0931]{Peng Jia}
\affiliation{College of Physics and Optoelectronics, Taiyuan University of Technology, Taiyuan, 030024, China}

\author{Jiaxin Li}
\affiliation{College of Physics and Optoelectronics, Taiyuan University of Technology, Taiyuan, 030024, China}

\author{Yu Song}
\affiliation{College of Physics and Optoelectronics, Taiyuan University of Technology, Taiyuan, 030024, China}

\author{Chengkun Hou}
\affiliation{College of Physics and Optoelectronics, Taiyuan University of Technology, Taiyuan, 030024, China}

\author{Yushan Li}
\affiliation{College of Physics and Optoelectronics, Taiyuan University of Technology, Taiyuan, 030024, China}



\begin{abstract}
The rapid advancement of image analysis methods in time-domain astronomy, particularly those leveraging artificial intelligence (AI) algorithms, has highlighted efficient image pre-processing as a critical bottleneck affecting algorithm performance. Image pre-processing, which involves standardizing images for training or deployment of various AI algorithms, encompasses essential steps such as image quality evaluation, alignment, stacking, background extraction, gray-scale transformation, cropping, source detection, astrometry, and photometry. Historically, these algorithms were developed independently by different research groups, primarily based on Central Processing Unit (CPU) architecture for small-scale data processing. This paper introduces a novel framework for image pre-processing that integrates key algorithms specifically modified for Graphics Processing Unit (GPU) architecture, enabling large-scale image pre-processing for different algorithms. To prepare for the new algorithm design paradigm in the AI era, we have implemented two operational modes in the framework for different application scenarios: Eager mode and Pipeline mode. The Eager mode facilitates real-time feedback and flexible adjustments, which could be used for parameter tuning and algorithm development. The pipeline mode is primarily designed for large scale data processing, which could be used for training or deploying of artificial intelligence models. We have tested the performance of our framework using simulated and real observation images. Results demonstrate that our framework significantly enhances image pre-processing speed while maintaining accuracy levels comparable to CPU based algorithms. To promote accessibility and ease of use, a Docker version of our framework is available for download in the PaperData Repository powered by China-VO, compatible with various AI algorithms developed for time-domain astronomy research.
\end{abstract}

\keywords{GPU Computing (1969) --- Astronomical Image Processing (2306) --- Time domain astronomy (2109)}


\section{Introduction} \label{sec:intro}
Time-domain astronomy focuses on studying celestial bodies and phenomena that evolve over time, revealing variations on minute scales and potentially leading to groundbreaking discoveries \citep{kaiser2004pan, rau2009exploring, udalski2015ogle, tonry2018atlas, ivezic2019lsst, bellm2019zwicky, kou2019optical, lokhorst2020wide, xu2020real, dyer2020gravitational, liu2021sitian, law2022low, beskin2023saint}. Data of time-domain astronomical event candidates are typically gathered using telescopes with survey mode and follow-up data of these candidates are often obtained by other specialized follow-up telescopes, which generally can only observe one celestial object at a time \citep{baranec2021automated, Jia2023AJ, jia2023simulation, scharwachter2024time}. The extensive deployment of sky survey telescopes for capturing observational images and identifying candidates for time-domain astronomical events across various observation modes leads to the detection of a significant number of potential time-domain astronomical events each night \citep{Coughlin_2023}. In contemporary time-domain astronomical observations, events displaying significant variations in magnitude or position can be detected utilizing an image subtraction data processing pipeline. This approach identifies optical transient candidates through image subtraction and subsequently classifies these transients using classification algorithms. Meanwhile, events characterized by minor magnitude variations can be effectively detected through high-accuracy photometry coupled with light curve classification algorithms. Typically, both of the above data processing modes are employed concurrently during the processing of real observation data to enhance reliability and precision. Given the substantial volume of observational data produced by time-domain astronomical surveys, alongside the demands for follow-up observations, efficient and timely data processing becomes imperative. Consequently, the primary challenge in time-domain studies arises from the juxtaposition of limited observational resources and the abundance of time-domain astronomical events that require investigation.\\

To enhance the efficiency in observing time-domain astronomical events, it is crucial to identify genuine events from a vast array of candidates as early as possible. For algorithms based on image subtraction, it is essential to achieve high completeness in detecting optical transient candidates, while accurately identifying true optical transients is equally important. In the context of light curve classification algorithms, the primary objective is to classify light curves with high precision. However, conventional light curve classification methods often analyze entire light curves, and results obtained by image subtraction methods can be significantly compromised by varying noise levels, resulting in missed critical opportunities for follow-up observations. Consequently, essential early-stage multicolor photometry results and spectral information pertaining to time-domain astronomical events are frequently overlooked, posing challenges for subsequent studies in this field. The resurgence of artificial intelligence (AI) in time-domain astronomy presents numerous possibilities. By emulating human behavior and taking advantage of vast datasets, advancements in algorithms, and the development of novel computational hardware \citep{sloan2023}, AI has transformed the landscape of time-domain astronomical observations. For straightforward tasks such as identifying optical transients or classifying light curves, machine learning algorithms, a subset of AI, have been extensively explored as potential substitutes for human input in devising effective classification features \citep{mahabal2008automated, duev2019real, jia2019optical, hinners2018machine, burhanudin2021light, sooknunan2021classification, yu2021survey, abraham2021machine, martinez2021method, demianenko2023understanding}. Meanwhile, another branch of AI, optimal search and reinforcement learning may replace human scientists in the selection of optimal parameters for data processing pipelines and the management of multiple telescopes engaged in time-domain observations \citep{landman2021self,jia2023observation}. The ultimate aspiration for AI is the development of Superintelligent AI that could exceed human capabilities in detection of optical transients with minimal data points, through the integration of data from various dimensions. Innovative approaches, such as multimodal neural networks, are being developed to amalgamate spectral data, images, and star catalogs for the classification of transients into distinct categories \citep{rehemtulla2024zwicky}. Furthermore, emerging methods that incorporate large language models (LLMs) and large vision models (LVMs) aim to capture more intricate details about celestial objects for classification purposes \citep{sotnikov2023language, shao2024astronomical, leung2024towards, fu2024versatile}. As a result, we can expect the emergence of increasingly sophisticated methodologies in the future.\\ 

Although extensive research has been conducted on various AI algorithms for time-domain astronomy, there is a lack of fundamental tools specifically designed for these algorithms. This paper concentrates on AI algorithms utilized in image processing. For image processing tasks, such as the classification of optical transients or photometry, it is essential to perform image pre-processing. However, the pre-processing algorithms developed for general-purpose image processing are often inadequate for astronomical applications. For instance, while image augmentation is crucial for classification algorithms like optical transient classification, the current image augmentation techniques are primarily built using OpenCV, which only supports images formatted with three channels and 8-bit $2^8$ grayscale values. However, astronomical images are often observed by single-frame, single-channel images with a larger dynamic range than standard images. In addition, these images often need to be aligned as image sequences or images with multiple colors for light curve classification or transient discovery. Consequently, extensive data pre-processing is crucial to process data to be AI-ready level, before applying AI algorithms effectively. Some key pre-processing steps may include:
\begin{itemize}
  \item \textbf{Noise reduction:} removing or mitigating noise within images to ensure accurate photometry and reliable light curves. Crucially, this process requires understanding and quantifying the underlying uncertainties so that they can be appropriately incorporated into the data products, leading to more robust photometric measurements and light curve analyses.
  \item \textbf{Image alignment:} aligning images captured at different epochs or at different observation bands, enabling the creation of deep coadded images through stacking or identification of transient candidates via image subtraction.
  \item \textbf{Image formatting:} structuring processed images for AI algorithms compatibility, such as compiling aligned images into data cubes or extracting light curves based on photometric data across multiple image frames.
\end{itemize}
Astronomers have developed various image pre-processing algorithms over an extended period, such as SWarp \citep{bertin2010swarp}, reproject \citep{deforest2004re}, SExtractor \citep{bertin1996sextractor}, Photutils \citep{larry_bradley_2024_10967176}, and HOTPANTS \citep{becker2015hotpants}. These are not ideal for the demands of modern AI algorithms and the hardware which AI algorithms are implemented in. In image pre-processing tasks, GPUs have significant advantages over CPUs. The GPU possess thousands of smaller and simpler processing units capable of handling thousands of threads simultaneously, making them particularly suited for tasks requiring numerous identical operations. Besides, GPUs utilize a Single Instruction Multiple Data (SIMD) architecture, which allows them to execute the same operation on multiple data points simultaneously. Their stream processors can process data streams in parallel, and their high-bandwidth VRAM can quickly read and write large amounts of data, reducing data transfer bottlenecks. Modern GPUs are also equipped with specialized hardware acceleration units, such as NVIDIA's CUDA cores and Tensor cores, which greatly enhance the efficiency of specific types of computations. Applying GPU acceleration technology to astronomical image pre-processing can significantly improve processing speed and provide more efficient solutions for handling large-scale datasets, thereby advancing the rapid development of astronomical research. Besides, developing astronomical image pre-processing algorithms under the GPU architecture could take better use of the GPU and reduce the requirement of the CPU. Therefore some algorithms have been adapted for GPUs in astronomical data processing \citep{zhao2013accelerating, li2013gpu}, but these algorithms have not been integrated together as a framework. To address this, a novel framework is needed. This framework should incorporate algorithms specifically designed for GPUs and offer seamless integration between them. \\

This paper introduces a framework for GPU accelerated image pre-processing specifically designed for AI algorithms. The framework integrates several crucial widely used image pre-processing algorithms together, including: image quality assessment, background estimation, image alignment, image subtraction, source detection, and gray scale transformation. To cater to varying application needs, our framework offers two distinct operational modes for different application requirements: eager mode and pipeline mode. We have tested our framework using simulated images, validating its effectiveness and uncertainties quantitatively. Furthermore, real observational data from the Ground based Wide Angle Camera (GWAC) is utilized to demonstrate the effectiveness of our algorithms in practical scenarios. The GWAC is a dedicated wide-field survey telescope array specifically designed to monitor transient events and time-varying phenomena, with the goal of providing timely detection and follow-up observations for rapidly changing astronomical events. Images used in this paper are obtained by telescopes with aperture of 18 cm and exposure time of 18 seconds. Each images has $4196\times 4136$ pixels and has a pixel scale of 11.7 arcsec. There are two data processing systems in the GWAC: an automated transient candidate validation system for fast optical transient detection, along with a light curve classification system for the analysis of transient events with small magnitude variations \citep{xu2020gwac}. Currently, our research group is in the process of developing an Agent based pipeline management system, alongside AI algorithms for transient detection and light curve classification. The framework proposed in this paper is used as basic tool, which will be seamless integrated into aforementioned two systems. The structure of the paper is laid out below. In Section \ref{sec:package}, we deliver an in-depth exploration of the two operational modes and the design of the pipeline. In Section~\ref{sec:Application}, we discuss the results derived from the framework testing, demonstrating its capability and precision with real observation data. In Section~\ref{sec:ConFut}, we make conclusions and anticipate our future works.\\

\section{Design of the Image Pre-processing Framework} \label{sec:package}
\subsection{Different Modes Design of the Framework} 
There are two stages in the application of AI algorithms in astronomy \citep{ansel2024pytorch}. The first involves algorithms development for specific tasks, with fixed input and output formats, which require specialized data augmentation and pre-processing methods. An example widely discussed is the transient classification algorithm \citep{jia2019optical, gomez2020classifying, turpin2020vetting, goode2022machine}. The second reflects the evolution of algorithms towards operational machine learning (ML-Ops), where models autonomously make real-time decisions \citep{gunny2022software}. Therefore, we have developed different per-processing modes for different applications.
\begin{itemize}
  \item For the development of task-specific AI algorithms, scientists require a data pre-processing framework that allows continuous function calls, known as the eager mode \citep{galeone2019hands}. In this mode, a variety of pre-processing algorithms—including image quality assessment, background estimation, image alignment, image subtraction, source detection, grayscale transformation, and visualization—are seamlessly incorporated into our framework package. These algorithms allow users to conduct data pre-processing efficiently by invoking functions within the framework. The key characteristics of this mode are: simplicity and intuitiveness, flexible parameter configuration, and quick result visualization. To meet requirements of the mode, we define functions separately according to their abilities and define the visulization part in the function. This approach is particularly beneficial for applications that require rapid verification of processing effects and improved workflows, such as preliminary analysis and processing of astronomical observation data.
  \item In the deployment stage of AI algorithms within ML-Ops frameworks, the data pre-processing framework is integrated together and run with predefined parameter sets. The parameter sets will be used as the default parameter sets unless some of them require updating. To address this requirement, we employ the pipeline mode in our framework. The pipeline mode is a data pre-processing approach designed for batch processing and model-specific tasks, implemented using the NVIDIA DALI (Data Loading Library). DALI, developed by NVIDIA, is a high-performance, open-source library for data loading and augmentation, compatible with mainstream deep learning frameworks \citep{nvidiadali}. Using the DALI framework, our pipeline can quickly load data and seamlessly integrate with deep learning models, thus enhancing the efficiency of the model-training process. In this mode, data pre-processing steps are predefined as a computation graph, where each node represents a specific data processing operation. Data is passed and processed according to the structure of this graph. The graph mode is particularly well-suited for large-scale batch processing tasks. The framework maximizes the use of GPU computing resources, intelligently allocating and releasing memory resources to avoid memory exhaustion or waste.
\end{itemize}

\subsection{Major Function Design in the Framework} \label{subsec:components}
In contemporary computer architecture, astronomical image pre-processing tasks are typically distributed between CPUs and GPUs. However, this approach can lead to significant time delays when images are processed separately by these components. The frequent transfer of image data between CPUs and GPUs via PCIe results in substantial overhead. To optimize performance, our framework adopts a GPU-centric approach, where all images are loaded into the GPU and all image pre-processing tasks are executed on the GPU. This strategy minimizes data transfer and leverages the parallel processing capabilities of GPU for improved efficiency. In the following subsections, we will explore the design of key functions within our framework, demonstrating how this GPU-focused approach enhances overall system performance.\\

\subsubsection{Image Quality Assessment Function} \label{subsubsec:imgqualitycheck}
\begin{figure} 
   \begin{center}
   \begin{tabular}{c} 
   \includegraphics[height=10cm]{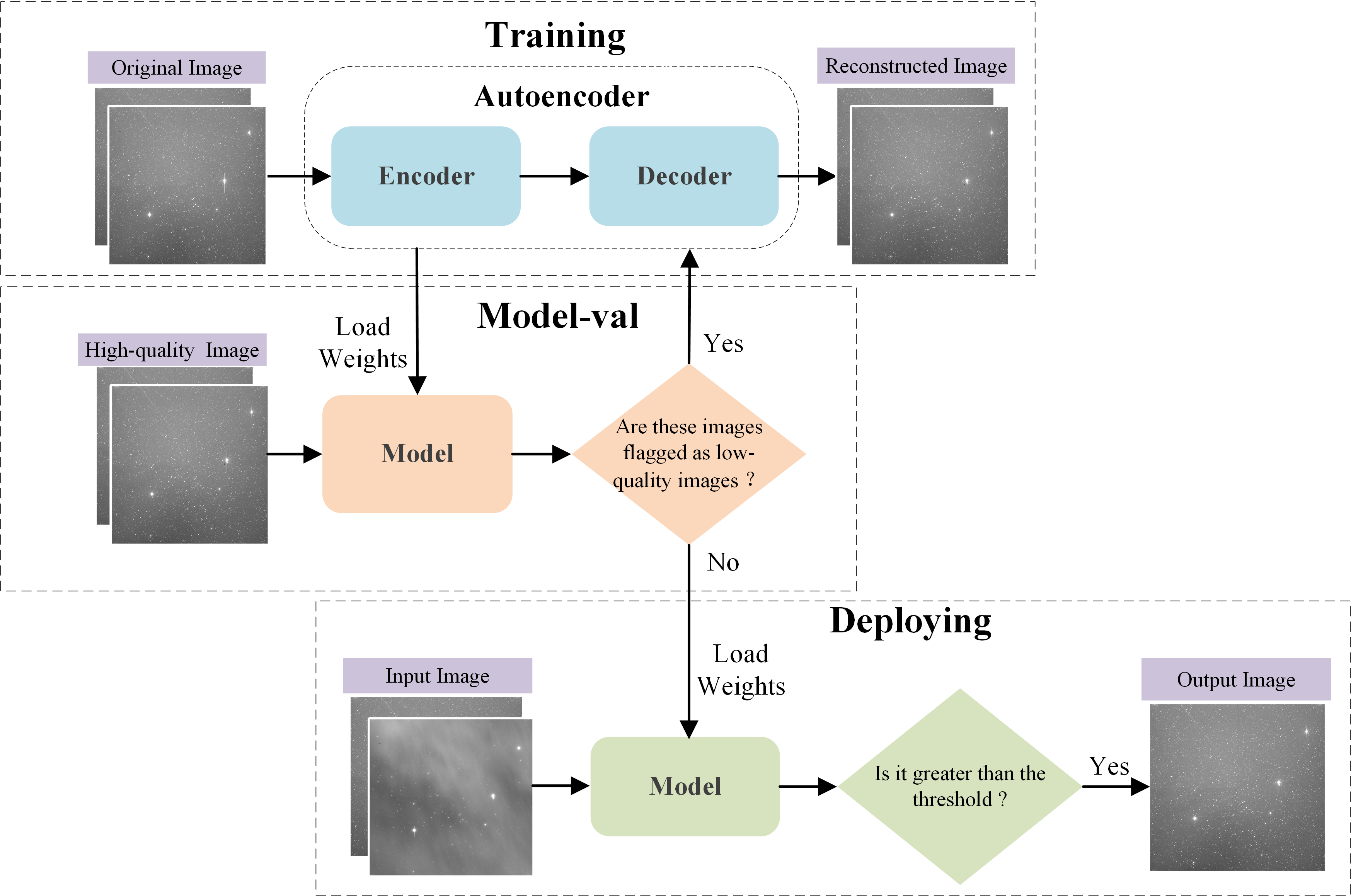}
	\end{tabular}
	\end{center}
   \caption
   { \label{fig:IQA} 
The process of training, validating, updating, and deploying the image quality assessment function. As shown in the figure, we first train an autoencoder using high-quality images, validate and update the model periodically, and then deploy the trained model to filter out high-quality images.}
   \end{figure}

The image quality assessment function serves as the first step in the astronomical image pre-processing procedure. Depending on the specific design, the image quality assessment function is capable of detecting medium to large-scale non-uniformities, and automatically picks up images with adequate quality or segments astronomical images into small patches on the basis of their quality. The assessment aids scientists in automatically selecting high-quality images while filtering out those images with large to middle scale non-uniformities, which are affected by technical issues like shutter failure or environmental factors such as moonlight and cloud cover. Implementing image quality assessment methods can significantly enhance photometric accuracy and reduce detection errors of optical transients. While traditional approaches often required substantial human effort to design appropriate features, our framework employs a deep learning based method proposed by \citet{jia2024image}. This innovative approach uses an autoencoder architecture, as introduced by \citet{wang2014generalized}, to learn features of high-quality images through self-supervised learning, as illustrated in Figure~\ref{fig:IQA}. The training process enables the encoder to effectively capture the characteristics of optimal images. After training, we apply the trained encoder to reconstruct observation images. By comparing these reconstructed images with their originals, we can assess the quality of observed images. Images exhibiting larger reconstruction errors indicate a greater deviation from high-quality standards, thus signifying relatively lower quality. \\ 

The image quality assessment function employs two distinct methods: Pre-training and Deployment. The Pre-training method trains the encoder on new data, enabling adaptive learning based on current astronomical conditions. The Deployment method uses pre-trained weights to select high-quality images based on predefined thresholds. To optimize computational efficiency, our framework assesses the computation capabilities and available memory of the GPU. For different modes, we use different strategies. In Eager Mode, we follow the original image quality assessment design, incorporating training, and deployment/validation processes. This approach enables robust and efficient image quality evaluation. In Pipeline Mode, the algorithm processes image batches in parallel, maximizing throughput. Besides, the algorithm with the pipeline mode integrates the GPU-optimized image quality assessment algorithm into the ML-Ops framework. A dedicated function, mode-val, periodically evaluates the performance of the image quality assessment algorithm using high-quality images. If the algorithm flags these images as unacceptable, it triggers an alert, prompting the acquisition of additional images for further training.\\

\subsubsection{Background Estimation and Removal Function}
This paper defines background noise as middle to large spatial scale gray scale variations arising from sources such as sky background noises, artificial light pollution, glow light from the detector and stray light emanating from proximate luminous objects. While Image Quality Assessment function can identify images degraded by such non-uniformities, accurate estimation and removal of these background artifacts is often crucial. This is because background noise not only influences the overall gray scale values but also introduces variations across intermediate spatial scales. Therefore, precise background estimation and removal function becomes paramount for both enhancing image quality and ensuring the reliability of subsequent photometric analyses \citep{Popowicz2015}. To address this issue, the background estimation and removal function employs standard iterative statistical methods to mitigate background variations \citep{Masias2013}. It is important to note that the primary purpose of our background estimation and removal function is to reduce false positives in transient detection and increase photometric accuracy. Although our method effectively removes background noise variations, there remains a possibility of inadvertently eliminating dim targets. It is a trade-off that we have to make during real data processing steps.\\

In the background estimation and removal function, the original image is segmented into smaller regions. This division can be based on either predefined masks generated by the Image Quality Assessment function or rectangular regions defined by the user. The default dimensions of these rectangular regions are $64 \times 64$ pixels and according to our experience, we need at least 100 pixels to keep statistical effective results. Each sub-region will be processed by a GPU thread. In each sub-regions, we will calculate the statistical values of gray scales of each pixel, defining parameters to eliminate outliers based on a predefined gray scale distribution. We offer four distinct methods for background noise removal: median estimation, mean estimation, $\sigma$ estimation, and mode estimation. For median and mean estimation, we calculate the median or mean gray scale values of the pixels within the sub-region. In the case of $\sigma$ estimation, we use the median and mean values alongside standard deviation. For the mode estimation, we uses a fixed weighting factor to integrate the median and mean gray scale values. All four methods are based on the implementation provided by the Photutils \citep{larry_bradley_2024_10967176} and they are defined in equation 1, 2, 3 and 4.

\begin{equation}
C_{\text{median}}  = \left\{
\begin{array}{ll}
I_{\frac{N+1}{2}},  & \text{if } N \text{ is odd} \\
\frac{I_{\frac{N}{2}} + I_{\frac{N}{2} + 1}}{2}, & \text{if } N \text{ is even}
\end{array}
\right.,\\
\end{equation}
\begin{equation}
C_{\text{mean}}= \frac{1}{N} \sum_{i=1}^{N} I_i, \\
\end{equation}
\begin{equation}
C_{\sigma} = 2.5 \times C_{\text{median}} - 1.5 \times C_{\text{mean}},\\
\end{equation}
\begin{equation}
C_{\text{mode}} = \text{median\_factor} \times C_{\text{mean}} - \text{mean\_factor} \times C_{\text{mean}}.
\end{equation}

Where $C_{\text{median}}$ is background noise obtained with median estimation method, $C_{\text{mean}}$ is background noise obtained with mean estimation method, $C_{\sigma}$ is background noise obtained with $\sigma$ estimation method, and $C_{\text{mode}}$ is background noise obtained with mode estimation method. $I_{\frac{N+1}{2}}$ and $\frac{I_{\frac{N}{2}} + I_{\frac{N}{2} + 1}}{2}$ are median value of pixels and $N$ is total number of pixels in the sub-region. $I_i$ is gray scale value for pixel $i$. $\text{median\_factor}$ and $\text{mean\_factor}$ are weighted factors for background noise estimation. With the above equations, we determine the upper and lower limits based on the calculated central value and the specified $\sigma$ range, exclude data outside these limits, and recalculate the central value and standard deviation. Once the maximum number of iterations is reached or the data stabilizes, the background value for each sub-region is obtained according to the predefined statistical method. Throughout this process, we leverage the parallel computing advantages of the GPU to simultaneously process all sub-regions. The calculations of central values, standard deviations, and data mask updates necessitate extensive memory access operations. The CUDA architecture optimizes these memory access patterns, minimizing thread contention and significantly enhancing operational performance.\\

To enhance the smoothness and continuity of the background estimation, we apply a median filter to the estimated background values. This process reduces noise and irregularities, resulting in a more uniform background representation. On the GPU, we implement this smoothing operation by executing multiple threads in parallel, with each thread processing distinct sections of the data. This parallel approach significantly accelerates the median filter computation. Following the smoothing process, we employ spline interpolation to scale the refined background values back to the original gray scale values. This step produces a background matrix that accurately represents the underlying noise structure while maintaining the original image's scale and resolution. The combination of parallel processing and interpolation techniques ensures both efficiency and precision in our background estimation procedure.\\

\subsubsection{Image Alignment Function}
While variations in PSF morphology stemming from observational conditions and diverse telescopes can be explained by optical imaging theory—using tools like Zernike polynomials and pupil functions to model wavefronts and calculate theoretical PSFs—positional shifts across different field of views present a greater challenge. These shifts, often caused by random misalignments or distortions, which could change between different observation epochs, cannot be easily derived through standard physical optics calculations. To solve this problem, it is crucial to use algorithms capable of accurately calibrating the positions of celestial objects within the image plane. Image alignment is a fundamental and long-established aspect of image pre-processing. It ensures that images captured at different times or by various instruments are properly aligned, thereby facilitating subsequent analyses. To optimize the image alignment process, our framework employs a GPU-accelerated function comprising three key steps: pixel coordinate mapping, affine matrix estimation, and image reprojection. The following parts will delve into the design and implementation of each step in the function.\\

\begin{figure} 
   \begin{center}
   \begin{tabular}{c} 
   \includegraphics[height=8cm]{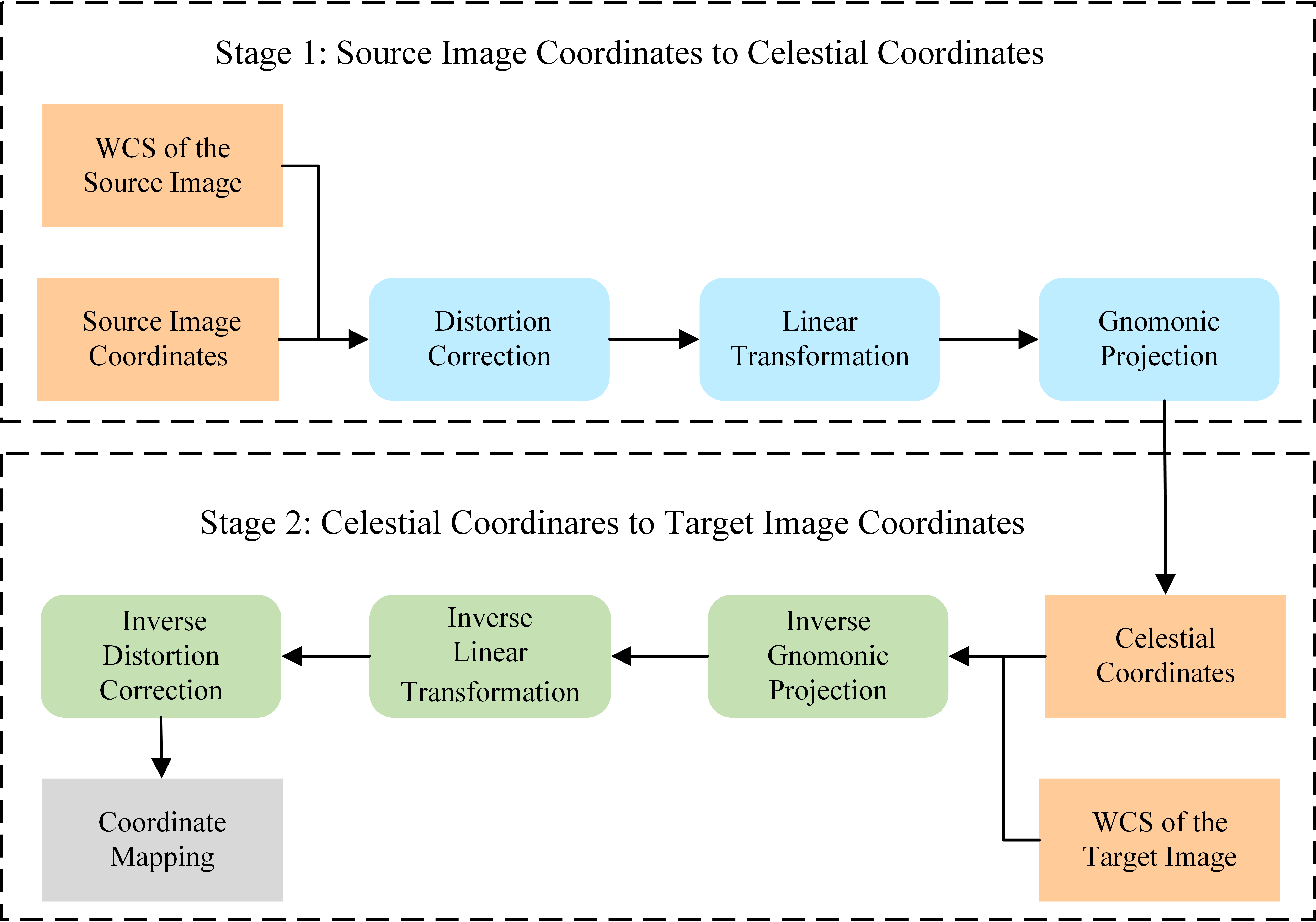}
	\end{tabular}
	\end{center}
   \caption
   { \label{fig:The coordinate mapping process} 
The process of obtaining the coordinate mapping. Since each pixel coordinate undergoes the same calculation, leveraging the advantages of GPU parallel computing can significantly reduce the time required for this process.}
   \end{figure}

We utilize the celestial coordinate system as a reference frame to map pixels from original images onto a unified coordinate system. This unified system employs a lower spatial resolution pixel grid, ensuring compliance with physical principles that prevent resolution increases during mapping. In practical applications, we allocate a GPU thread to each pixel, performing coordinate transformations to determine corresponding positions in the target image. Furthermore, we implement distortion correction using a specific equation, leveraging source image header information to model and compensate for nonlinear distortions introduced by the optical system through polynomial transformations. Given that GPUs typically feature thousands of threads per block and multiple blocks, we can process a large number of pixels simultaneously in real applications. Both coordinate mapping and distortion correction are executed on the GPU, significantly reducing computation time. The computation procedure is shown in figure \ref{fig:The coordinate mapping process} and we will discuss details below.\\
As discussed above, coordinate mapping is divided into two stages: conversion from source image coordinates to celestial coordinates and conversion from celestial coordinates to target image coordinates. The conversion from source image coordinates to celestial coordinates involves three steps: distortion correction, linear transformation, and gnomonic projection. Specifically, we first extract the World Coordinate System information from the source image header and use the polynomial distortion model coefficients to perform distortion correction on the pixel coordinates of the source image, as shown in equation~\ref{eq1},
\begin{equation} \label{eq1}
\left\{
\begin{array}{l}
\xi = x + \sum_{i=0}^{A_{ORDER}} \sum_{j=0}^{A_{ORDER} - i} A_{ij} \cdot x^i \cdot y^j\\
\eta = y + \sum_{i=0}^{B_{ORDER}} \sum_{j=0}^{B_{ORDER} - i} B_{ij} \cdot x^i \cdot y^j ,
\end{array}
\right.
\end{equation}
where $x$ and $y$ are original position of each pixel and $A_{i,j}$ and $B_{i,j}$ are parameters for coordinate transformation, $A_{ORDER}$ and $B_{ORDER}$ are maximal order of polynomials used in the function \citep{greisen2002representations}, which is inspired by \citet{alard1998method, bertin2006automatic}. After completing the correction, we apply the Coordinate Description Matrix from the image header for a linear transformation, using Equation \ref{eq2} to convert all pixel coordinates $(\xi, \eta)$ to an intermediate coordinate system $(u,v)$,
\begin{equation} \label{eq2}
\left( 
\begin{array}{c}
u \\ 
v 
\end{array} 
\right)
= 
\left( 
\begin{array}{cc}
\text{CD}_{1,1} & \text{CD}_{1,2} \\ 
\text{CD}_{2,1} & \text{CD}_{2,2} 
\end{array} 
\right)
\left( 
\begin{array}{c}
\xi - \text{CRPIX}_1 \\ 
\eta - \text{CRPIX}_2 
\end{array} 
\right).
\end{equation}

Subsequently, the gnomonic projection and Equation \ref{eq3} are used to transform the intermediate coordinates (u,v) into celestial coordinates (RA,Dec).

\begin{equation} \label{eq3}
\left\{
\begin{array}{l}
\text{RA} = \text{CRVAL}_1 + \arctan\left(\frac{u}{\cos(\text{CRVAL}_2) - v \sin(\text{CRVAL}_2)}\right) \\
\text{Dec} = \arctan\left(\frac{v \cos(\text{CRVAL}_2) + \sin(\text{CRVAL}_2)}{\sqrt{u^2 + (\cos(\text{CRVAL}_2) - v \sin(\text{CRVAL}_2))^2}}\right) 
\end{array}
\right.
\end{equation}

To obtain the pixel coordinates of these astronomical coordinates in the target image, we extract the World Coordinate System information from the target image header and perform an inverse gnomonic projection, inverse linear transformation, and inverse distortion correction on the $(\text{RA}, \text{Dec})$ to obtain the coordinates $(x', y')$ of the pixels from the source image in the coordinate system of the target image. This process achieves precise coordinate mapping between the two images. During this process, the efficient parallel scheduling mechanism of the SM is used to parallelize the computations across all threads, significantly improving the speed of coordinate transformation. It is worth noting that this process relies heavily on accurate WCS information. If the WCS information contains significant errors, the geometric transformation between the source image and the target image will be incorrect, resulting in misalignment of the same target across different epochs or bands, which can further affect subsequent data analysis. This issue is particularly pronounced in the edge of the field of views, where inaccuracies in distortion correction parameters cause larger positional offsets for targets. In crowded fields, where targets are densely distributed, WCS errors can exacerbate source confusion, significantly increasing the risk of false positives and missed detections. Additionally, due to the small separations between targets, even slight misalignment can lead to photometric contamination, thereby affecting the accuracy of subsequent target detection and photometry. Therefore, we recommend users to select non-uniform segmentation of the image plane, such as the method proposed in \citet{Jia2018}, and pay much attention to edge of the field of views.\\

Next, we implement the estimation of the affine matrix using OpenCV's estimate AffinePartial2D method in GPU version, and then perform image reprojection using the obtained affine matrix. Specifically, we use the affine matrix to obtain the coordinate mapping between the source image and the resulting image. For each pixel in the source image, the pixel value in the resulting image is calculated through bilinear interpolation based on the four neighboring pixels. By utilizing CuPy to perform matrix and interpolation operations on the GPU, we achieve high pixel value accuracy and increased speed during the reprojection process.\\

\subsubsection{Image Subtraction Function}
One of the primary scientific goals in time-domain astronomy is to identify optical transients, which exhibit significant changes in position or brightness over time. To efficiently detect these objects, it is essential to perform image subtraction on observation images made at different times. This method removes stars with constant magnitudes, isolating those with variable magnitudes or positions as optical transient candidates requiring further examination. The image subtraction function is very efficient for images of dense star fields. A prevalent technique for this purpose is the High Order Transform of PSF And Template Subtraction (HOTPANTS). HOTPANTS is an optimized image subtraction algorithm that focuses on determining an appropriate convolution kernel, enabling images captured under varying observational conditions to align as precisely as possible \citep{becker2015hotpants}. Since the introduction of HOTPANTS, the field of difference image analysis has seen the development of several alternative methods, such as DIAPL \citep{wozniak2000difference}, DANDIA \citep{bramich2008new}, and pyDIA \citep{Albrow2017}, as well as the more recent PyTorchDIA \citep{hitchcock2021pytorchdia}, which leverages GPU acceleration and automatic differentiation optimization. In this study, we have selected HOTPANTS as the core method, primarily due to its efficiency and stability in processing large field-of-view survey data, further enhancing its computational performance through GPU acceleration. Unlike PyTorchDIA, which relies on automatic differentiation optimization, our approach is based on the traditional image subtraction framework and optimizes key steps such as convolution kernel estimation, image convolution, and subtraction, avoiding additional computational overhead. This makes it better suited for real-time processing of large-scale survey data, such as transient object detection. While PyTorchDIA has notable advantages in complex noise modeling, our method prioritizes efficiency and stability in large-image processing, making it more suitable for wide-field survey observations and highly effective when handling ultra-large-scale datasets.\\

HOTPANTS employs a two-step process to generate difference images. First, it divides the input image (I) and template image (T) into a predefined number of stamps with the same pixel scale. Within each stamp, brighter sources are identified, and their pixel values are recorded. Only sources with pixel values significantly exceeding the background level are retained as candidate sources. These candidate sources are then used to calculate convolution kernel coefficients ($a_n(x, y)$) and the background difference ($B(x, y)$) related to their image positions. These values are subsequently used to construct the final convolution kernel ($K(u, v)$), as defined in equation~\ref{k(u,v)},
\begin{equation} \label{k(u,v)}
K(u, v) = \sum_{n=1}^{N} a_n(x, y) K_n (u,v),
\end{equation}
where $K_n (u,v)$ is the functional form of function $K(u, v)$. In this paper, it is a Gaussian profile modulated by the spatial factors within the convolution kernel as shown in equation~\ref{kn(u,v)},
\begin{equation} \label{kn(u,v)}
K_n(u, v) = e^{- \frac{u^2 + v^2}{2\sigma_k^2}} u^i v^j,
\end{equation}
where \( u \) and \( v \) are the coordinates within the convolution kernel, and \( \sigma_k \) represents the width of the chosen Gaussian profile. Multiple Gaussians with different \( \sigma_k \) can be chosen, each with a fixed spatial modulation order. The values of \( i \) and \( j \) and their sum do not exceed this order.\\
Subsequently, we apply a convolution operation to the template image to mitigate the effects of varying PSFs and background levels, as shown in equation~\ref{eq6},
\begin{equation} \label{eq6}
R(x, y) \otimes K(u, v) = I(x, y) + B(x, y).
\end{equation}

 \begin{figure} 
   \begin{center}
   \begin{tabular}{c} 
   \includegraphics[height=8cm]{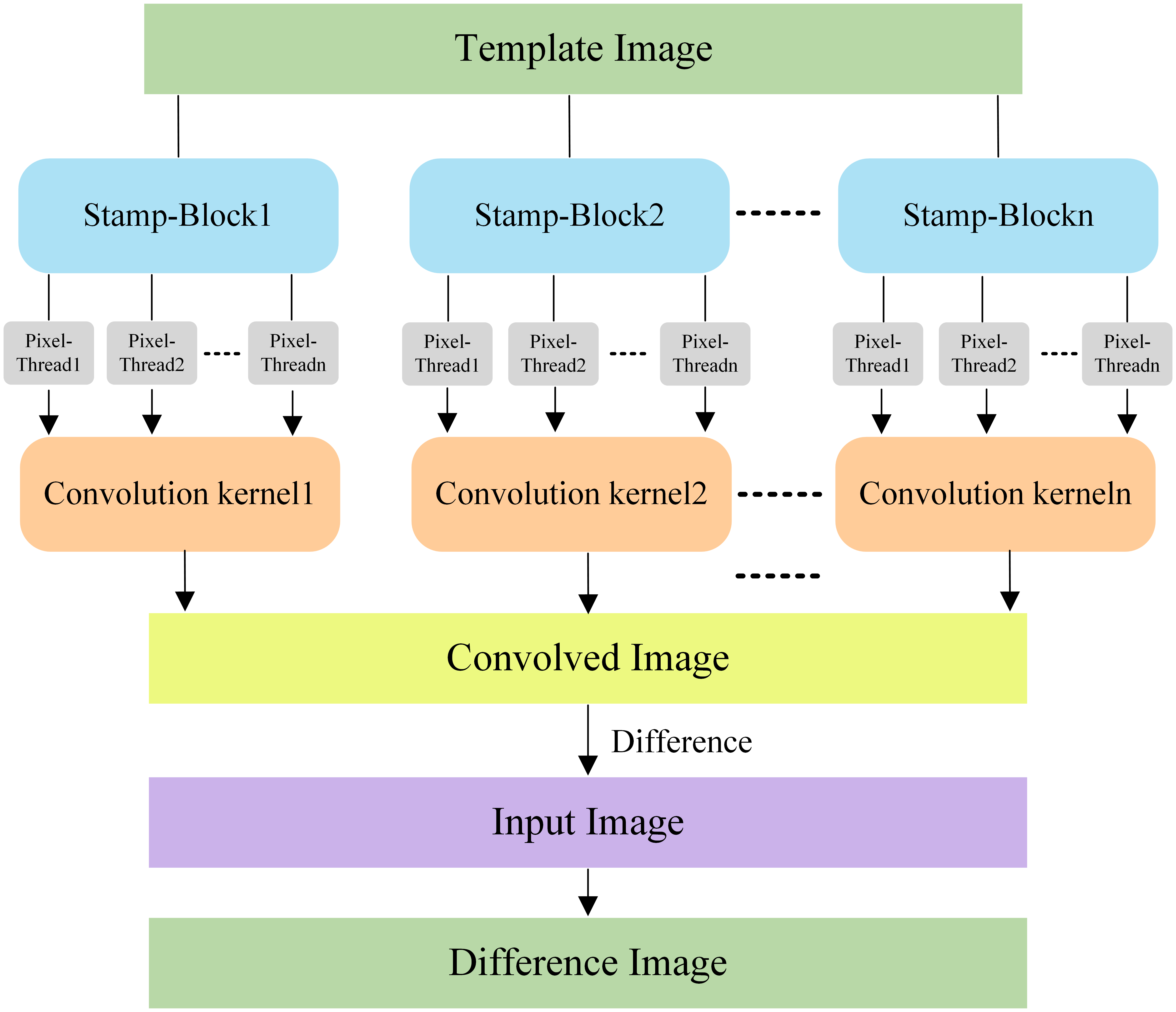}
	\end{tabular}
	\end{center}
   \caption
   { \label{fig: parallel convolution} 
The process of performing parallel convolution on the template image and generating the difference image. By leveraging the parallel computing power of the GPU, the time required for convolution can be significantly reduced.}
   \end{figure} 

This process is the most computationally intensive step in the entire algorithm, and leveraging GPU parallel computing can significantly boost efficiency, as discussed in \citet{Li2014}. The overall computation graph is shown in figure \ref{fig: parallel convolution}. To optimize this step, we begin by loading the data to be processed from global memory into shared memory, which improves data processing speed. We then allocate a thread block for each stamp of the convolution kernel size; by default, the size of the stamp is $21 \times 21$ pixels. The convolution computation is performed within shared memory, with each thread processing a single pixel in the stamp, utilizing the preloaded convolution kernel for the convolution operation. This approach takes advantage of the parallel processing capabilities of the GPU, allowing for simultaneous computation across multiple pixels. Once the convolution is completed, the results are written from shared memory to global memory. Finally, by subtracting the convolved template image from the input image, we obtain the difference image. This optimized process harnesses the power of GPU parallel computing to significantly reduce the computational time required for this critical step in the algorithm.

\subsubsection{Source Detection Function}
Time-domain astronomy encompasses another critical scientific objective: the classification of astronomical events according to their light curves, particularly those exhibiting subtle variations. The light curve classification process relies on source detection algorithm, which operates by identifying positional coordinates and measuring flux across clusters of adjacent pixels that exceed specified threshold values. Our framework solely focuses on identifying point like targets and measuring their flux. Since source detection process focuses on adjacent pixels, it will benefit from parallel computation. We have thus optimized traditional source detection and photometry algorithms using GPU parallel computing. By harnessing the powerful capabilities of GPUs, we could significantly reduce the time needed for source detection, enhancing processing speed while maintaining the accuracy and robustness of measurements.\\

\begin{figure}
   \begin{center}
   \begin{tabular}{c} 
   \includegraphics[height=8cm]{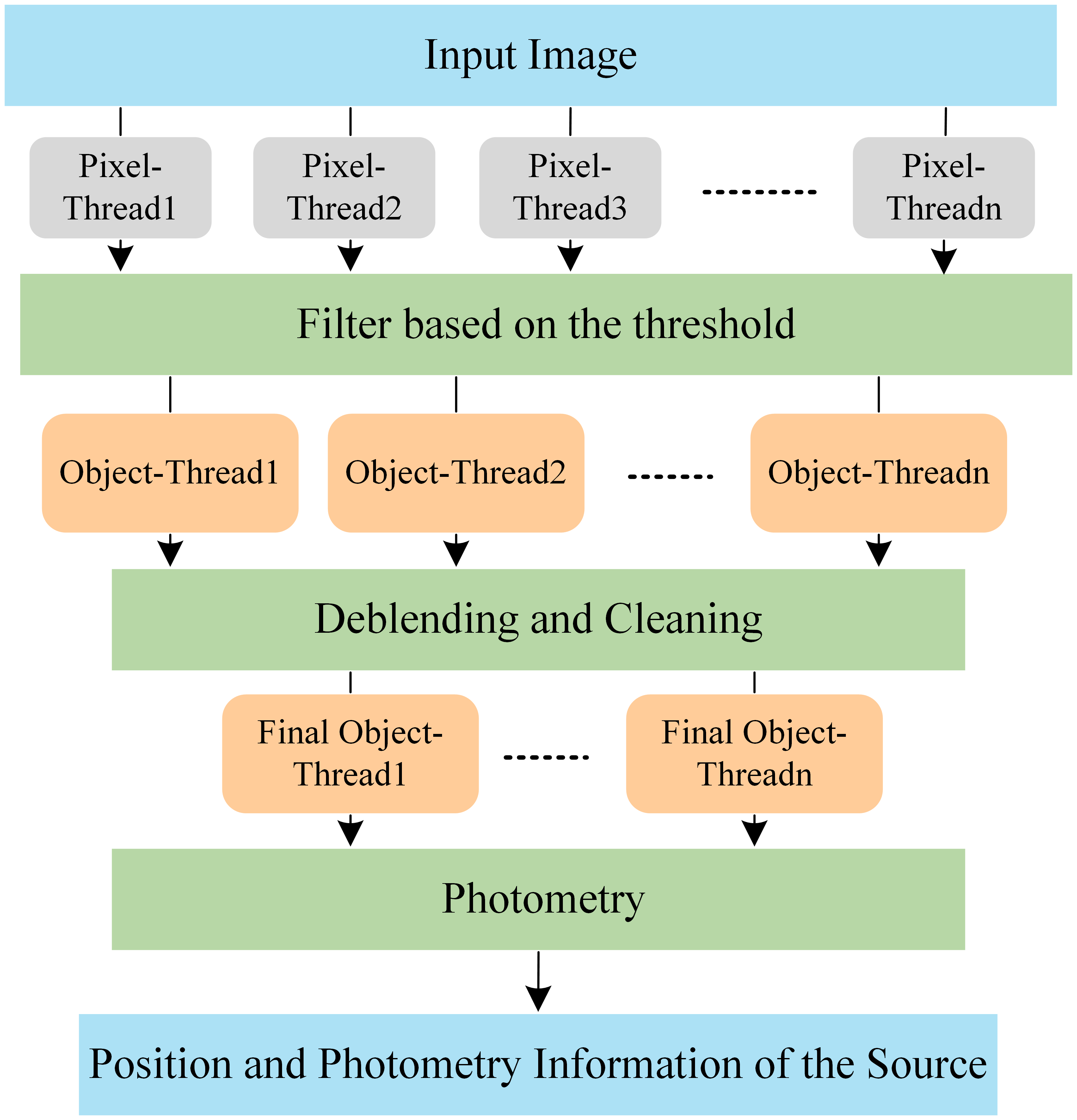}
	\end{tabular}
	\end{center}
   \caption
   { \label{fig: parallel SEX} 
The flowchart of the source extraction algorithm based on GPU parallel computing. By leveraging the parallel computing power of the GPU, the time spent on the four stages—detection, decomposition, cleaning, and photometry—has been significantly reduced.}
   \end{figure}

The source detection algorithm adapted in this paper is based on the approach proposed in SExtractor \citep{bertin1996sextractor}. This algorithm comprises four key stages: pixel detection, source deblending, cleaning, and photometry as shown in figure \ref{fig: parallel SEX}. During the pixel detection phase, we conduct a comprehensive scan of the entire image to identify pixels whose flux exceeds a specified main threshold. These pixels are then connected to form source candidates. This process leverages GPU parallel computing, with each thread assigned to process an individual pixel, ensuring swift and efficient pixel detection across the image. Following the pixel detection step, we perform a parallel pre-analysis of each source candidate. This analysis utilizes the Thrust library \citep{bell2012thrust} to calculate peak values and total flux for each candidate source.\\

Next, we implement the deblending phase to differentiate between nearby stars that might have been initially identified as a single target due to the low detection threshold. This process is optimized for parallel computation, with each source candidate assigned to a dedicated GPU thread. These threads efficiently read the peak value of their respective targets and calculate multi-level deblending thresholds. At each threshold level, the algorithm re-evaluates objects and splits them as necessary, constructing a binary tree structure rooted in the original source candidate. This hierarchical approach allows for a nuanced analysis of complex stellar configurations. The binary tree is then traversed from bottom to top, with each node processed by a separate thread, maximizing computational efficiency. To ensure the quality and relevance of detected sources, we apply a pruning mechanism to the binary tree. This pruning is based on rigorous intensity and spatial criteria, effectively filtering out spurious detection results and retaining only source candidates with statistically significant values.\\

In the cleaning phase, we eliminate false detection results that may have arisen due to noise or other artifacts in the image. This stage leverages parallel processing capabilities, with each GPU thread assigned to evaluate a specific object and its surrounding neighbors. The threads perform a detailed analysis to determine if the current object can be merged with any adjacent objects based on predefined criteria. If the analysis indicates that merging is appropriate, the current object is flagged as a false detection. This flagging process is carried out efficiently across multiple objects simultaneously. Following the identification of false detections, we execute a parallel operation to remove the marked pixels associated with these spurious objects.
By utilizing parallel processing, we maintain high computational efficiency while ensuring the integrity of our detected source catalog.\\

Following the previously described phases, we successfully identify celestial objects and their positions. The final stage of the source detection function involves photometry, where we calculate the flux of each object using several established methods, such as: aperture photometry, isophotal photometry and  auto photometry. This photometric analysis is particularly well-suited for parallel computation, as it is performed independently for image of each star. In our implementation, we have adapted the photometry procedures to leverage GPU capabilities. Each GPU thread is assigned to measure a single object, enabling the simultaneous processing of multiple celestial bodies. This parallelization significantly enhances the efficiency of our approach, resulting in rapid and accurate attribute measurements of each star. This approach allows us to handle large-scale astronomical datasets with improved speed and reliability, contributing to more efficient celestial object characterization in time-domain astronomy.\\

\subsubsection{Gray Scale Transformation Function}
Gray scale transformations play a crucial role not only in enhancing the visualization of astronomical image, but also increasing the neural network based source detection algorithm. Although utilizing unprocessed gray scale values is theoretically optimal for artificial intelligence algorithms, particularly in object classification and detection tasks, empirical evidence suggests that even elementary gray scale transformations can substantially improve algorithmic performance \citep{jia2022detection, sun2024artificial}. According to our experience, these transformation could significantly reduce the time cost in training of deep neural networks and increase the generalization ability of deep neural networks \citep{li2024csst}. This phenomenon can be attributed to several underlying mechanisms. While deep neural networks possess the inherent capability to learn such transformations internally, the process of acquiring these transformations typically demands significant computational resources and training duration, while simultaneously introducing superfluous parameters into the network architecture. To address this, we have developed a gray scale transformation function within our framework, optimized for GPU architecture. This function maps original pixel values to new gray scale values through diverse mathematical functions or histogram equalization methods. By implementing these gray scale transformation techniques on the GPU, we could achieve a several-fold increase in processing speed. The following sections will detail the main gray scale transformation methods and their GPU implementations, highlighting how these optimizations contribute to more efficient image processing in AI-driven astronomical applications. \\

Our framework includes several gray scale transformation methods, such as histogram equalization, Zsacle transformation sed in SAOimage \citep{joye2003new}, linear and non-linear transformations. These methods benefit from GPU based parallel computation as they operate on a pixel-by-pixel basis. Histogram equalization works by redistributing the gray scale values to achieve a more uniform distribution across the image. We harness GPUs to parallelize the tasks of counting pixels for each gray scale level, computing the gray scale histogram or cumulative distribution function, and creating a uniformly distributed target array. The interpolation function defined in CuPy is then used to map gray scale values of the original image to new ones \citep{nishino2017cupy}, effectively achieving equalization. For linear and non-linear transformations, including logarithmic, exponential, and power-law transformations, we rewrite the functions using CuPy. Each pixel is processed by a separate thread, which significantly reduces computation time. \\

\section{Application and Evaluation of the Pre-processing Framework}
\label{sec:Application}
\subsection{Introduction of the Application Scenario}
\begin{figure} 
   \begin{center}
   \begin{tabular}{c} 
   \includegraphics[height=7.8cm]{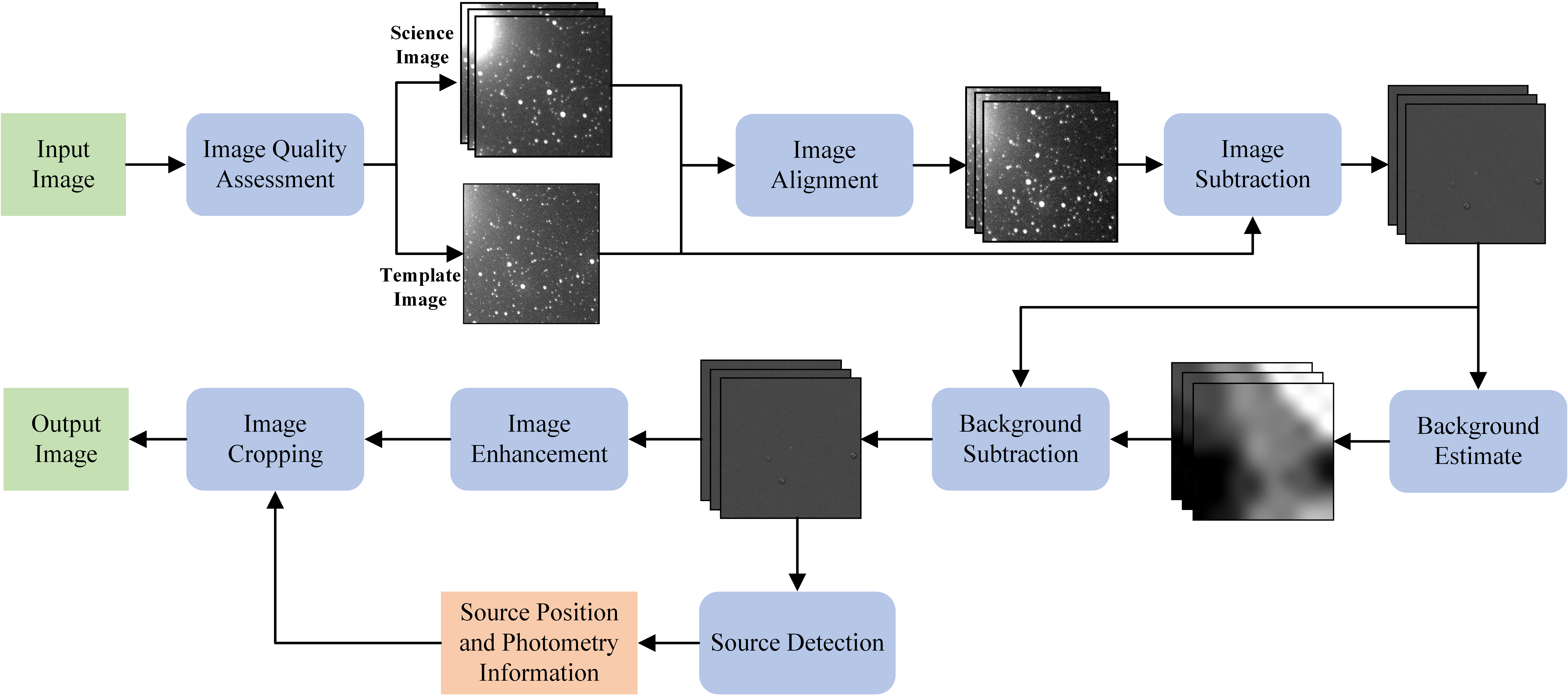}
	\end{tabular}
	\end{center}
   \caption
   { \label{fig:PipelineStructure} The diagram of the pipeline designed to carry out data pre-processing for the GWAC.}
   \end{figure}

The performance evaluation of our framework encompasses both simulated and real observational data. The real data is sourced from the GWAC, a network of robotic, multi-aperture optical telescopes. Initially designed to detect optical counterparts of gamma-ray bursts in conjunction with the Space Variable Object Monitor (SVOM), the GWAC's extensive coverage of over 5000 square degrees with 18 cm aperture telescopes enables it to serve broader astronomical purposes beyond high-energy transient detection. Two specialized data processing pipelines have been implemented to achieve these objectives: one utilizing image subtraction algorithms for rapid optical transient identification and subsequent follow-up observations \citep{bernardini2021svom}, and another employing light curve classification algorithms for time-domain astronomical event detection. Both pipelines demand swift and accurate classification capabilities to facilitate further scientific investigation. The integration of cutting-edge AI algorithms \citep{turpin2020vetting, chen2023meteor}, along with recently developed tools such as the data processing pipeline Supervisor and the multimodal transient classification neural network (Peng et al., 2025, submitted), promises to enhance real-time processing capabilities and overall efficiency. Given that these AI algorithms, particularly agent-based systems, frequently utilize various basic data pre-processing functions, the development of our GPU-based data pre-processing package represents a crucial advancement. In this section, we evaluate the performance of these GPU-based packages, which not only support AI model training and deployment but also seamlessly integrate into existing pipeline frameworks. A comprehensive illustration of how the proposed algorithms are integrated into the existing GWAC pipeline is presented in Figure~\ref{fig:PipelineStructure}.\\

\label{sec:test}

The integration of our algorithms into the existing data processing pipeline follows a systematic procedure. We begin with an image quality assessment function to identify high-quality images for use as templates and to mask low-quality images for further processing. Images with low quality affecting over 70\% of their area are removed from the pipeline to maintain data integrity. We select the highest quality image with smallest FWHM of PSF as a reference and estimate background noise with the background estimation function. We will iterate to remove background noises with the background estimation and removal function and utilize the image alignment function to align all other images to reference images. Following alignment, we employ the image subtraction function to obtain optical transients for the optical transient detection pipeline. Meanwhile, we carry out target detection and photometry from processed observation images with the SExtractor to obtain the light curve of each target \citep{feng2017real, bai2023photometric}. Additionally, we use gray scale transformation and visualization tools to provide data support for subsequent human inspection and deep learning based algorithms \citep{xu2020real, li2023magnetic, li2024white}. We will evaluate the performance of both classical methods and our pre-processing framework in executing these steps. Two evaluation criterion will be used for performance evaluation. We use the photometric accuracy to evaluate the photometry results. Meanwhile, we use the precision rate and recall rate to evaluate the detection results, where they are defined as Equation \ref{eq:recall}, \ref{eq:precision} and \ref{eq:F1},

\begin{equation}\label{eq:recall}
\text{Recall} = \frac{\text{True Positives}}{\text{True Positives} + \text{False Negatives}},
\end{equation}
\begin{equation}\label{eq:precision}
\text{Precision} = \frac{\text{True Positives}}{\text{True Positives} + \text{True Negatives}},
\end{equation}
\begin{equation}\label{eq:F1}
\text{F1} = \frac{2 \times (\text{Recall}+\text{Precision})}{\text{Recall} + \text{Precision}}.
\end{equation}

True Positives represent the number of genuine targets successfully identified by the detection algorithm, while False Negatives indicate the number of actual targets that the algorithm failed to detect. True Negatives denote the number of false targets incorrectly classified as genuine detections by the algorithm. The following sections will provide a detailed discussion of the performance metrics and time costs associated with each stage of the process.\\
   
Since we could not obtain the ground-truth values of real observation data and control noise levels, we can better qualitatively evaluate the precision and uncertainties of our algorithms. Therefore, we also use simulated images generated with the \textit{SkyMaker} to test the performance of our algorithms. The parameters for the simulated images are defined according to the parameters of the GWAC, which include the aperture size of the diameter, the field of view and the pixel size. Besides these parameters, we modify parameters, such as offset levels, background levels, density of celestial objects and seeing disc according to our requirements. We test our framework in a computer equipped with an Intel(R) Xeon(R) Gold 5218 CPU @ 2.30GHz and an NVIDIA GeForce RTX 3090 GPU. The CPU features 20 cores and 40 threads, while the GPU boasts 10496 CUDA cores and supports compute capability 7.5. Our system runs on Ubuntu 22.04.4 LTS, utilizing gcc and g++ compiler versions 11.4.0, Python version 3.10.4, and CUDA toolkit version 11.8. Key software packages used in our experiments include CuPy (version 13.3.0, CUDA 11.x), Photutils (version 1.8.0), NumPy (version 2.1.2), PyTorch (version 2.5.1), and Astropy (version 6.1.4). For comparison, we also utilize SExtractor (version 2.28.0), Hotpants (version 5.1.11), and SWarp (version 2.41.5). Given the challenge of directly comparing CPU and GPU computational capabilities, we selected components of similar price, assuming this reflects comparable performance potential.\\

\subsection{Performance Evaluation Using Simulated Images}

In this section, we perform a quantitative assessment of the pipeline's performance using simulated data. The data is produced by \textit{SkyMaker} based on the hardware specifications of the GWAC, as well as differing observation parameters. We will examine how our algorithms perform across various observation conditions, focusing on three specific scenarios: images with different positional shifts, varying levels of background, and differing sizes and densities of sources.

\subsubsection{Performance Evaluation with Different Offsets}
\begin{figure} 
   \begin{center}
   \begin{tabular}{c} 
   \includegraphics[height=6cm]{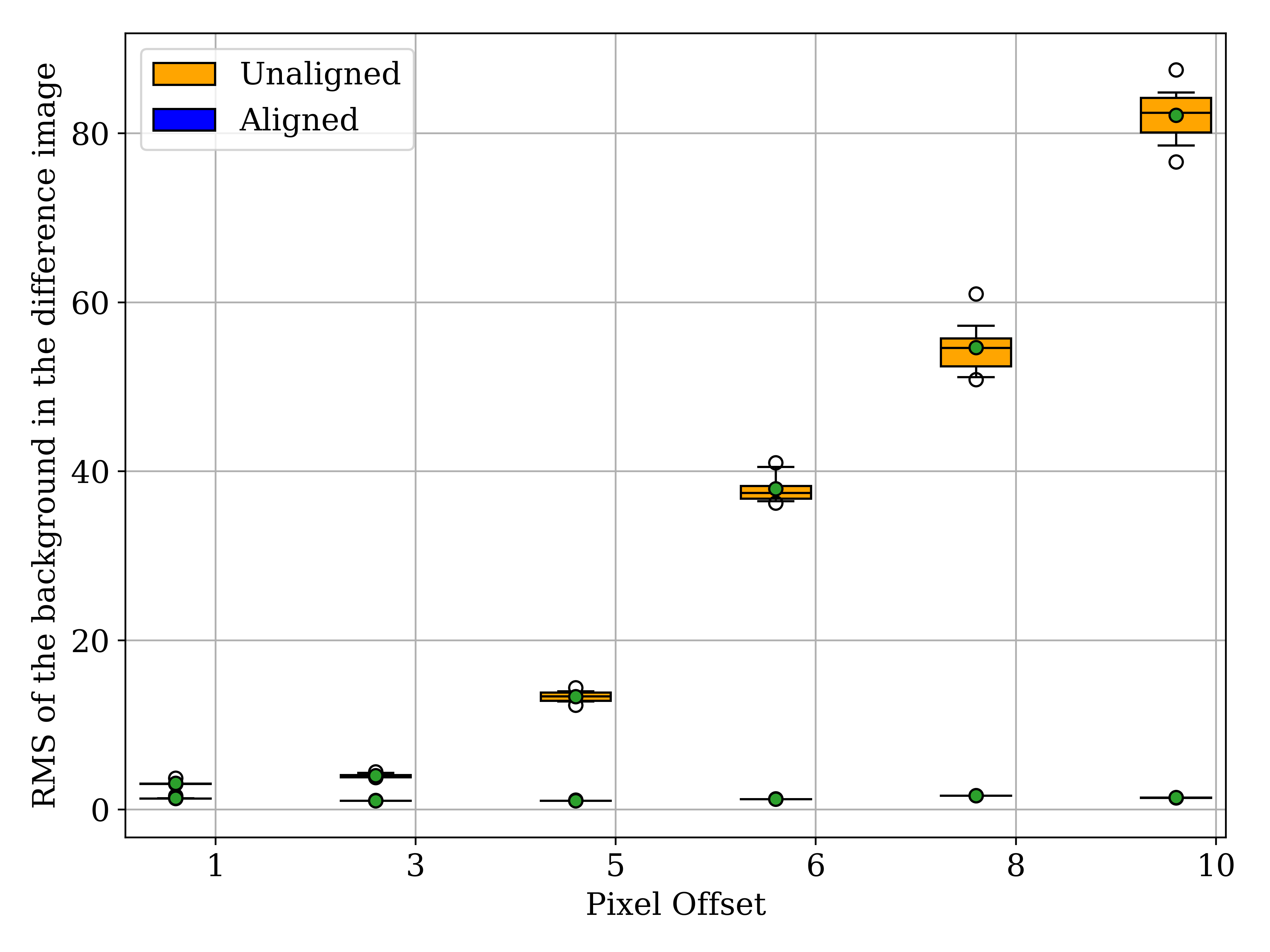}
	\end{tabular}
	\end{center}
   \caption
   { \label{fig:sim_pos_test} The figure demonstrates the effectiveness of our image alignment function by comparing residual between template and observation images under different offset levels. Compared to the unaligned case, the RMS of the background is significantly reduced by after alignment, achieving better subtraction results.}
   \end{figure}

We first examine realistic observational scenarios involving spatial offsets between different epochs, a common occurrence in time-domain astronomy where random deviations exist between intended and actual pointing directions. To simulate these conditions, we generate observational images with various positional offsets by adjusting the central \textit{RA} and \textit{Dec} coordinates of the simulated frames. For this analysis, we utilize SkyMaker to create 10 distinct star catalogs containing randomly distributed celestial objects. These catalogs are populated with stars having a magnitude limit of 15, a magnitude standard deviation of 3, and a spatial density of 100 sources per square degree. Using these catalogs as a foundation, we produce simulated images and introduce varying positional offsets by modifying their central coordinates, thereby mimicking the pointing uncertainties typically encountered due to telescope tracking errors or field drift. The effectiveness of our alignment procedure is evaluated by performing image subtraction between offset images (both pre- and post-alignment) and a reference frame, followed by an analysis of the residual background levels. As illustrated in Figure \ref{fig:sim_pos_test}, with our image alignment algorithm, we could effectively reduce residual background noise that  typically arises from positional offsets, with RMS smaller than. While unaligned images show a marked increase in error range with increasing offset, the aligned cases maintain consistently lower and more stable error values. With the low residual error, we could anticipate better optical transient detection results in processing real observation data.

\label{sim_pos:test}

\subsubsection{Performance Evaluation with Different Background Noise}
\begin{figure}
    \centering
    \begin{minipage}[t]{0.43\linewidth}  
        \centering
        \includegraphics[height=0.7\linewidth]{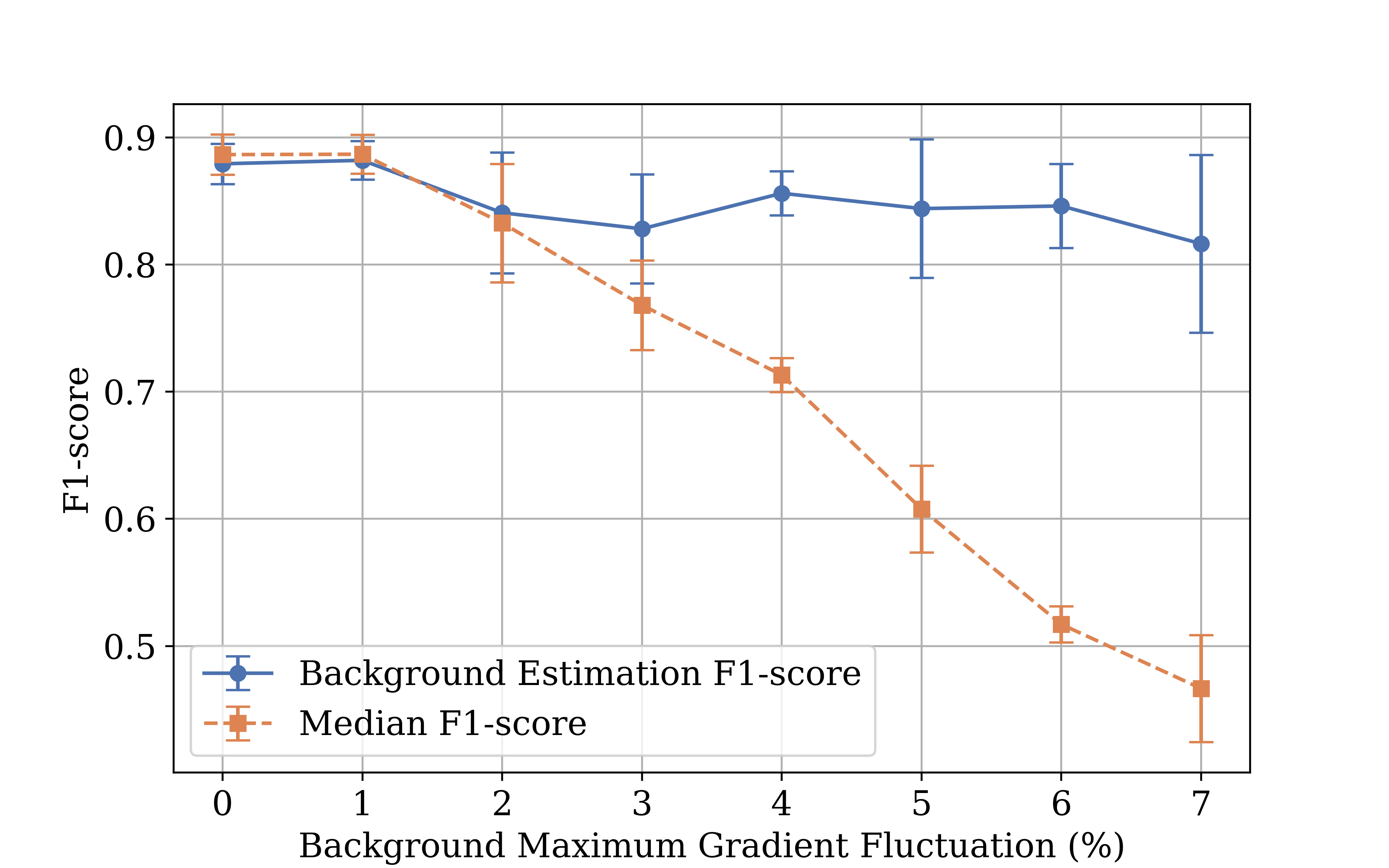}
        \label{fig:sim_recall}
    \end{minipage}
    \begin{minipage}[t]{0.43\linewidth}
        \centering
        \includegraphics[height=0.7\linewidth]{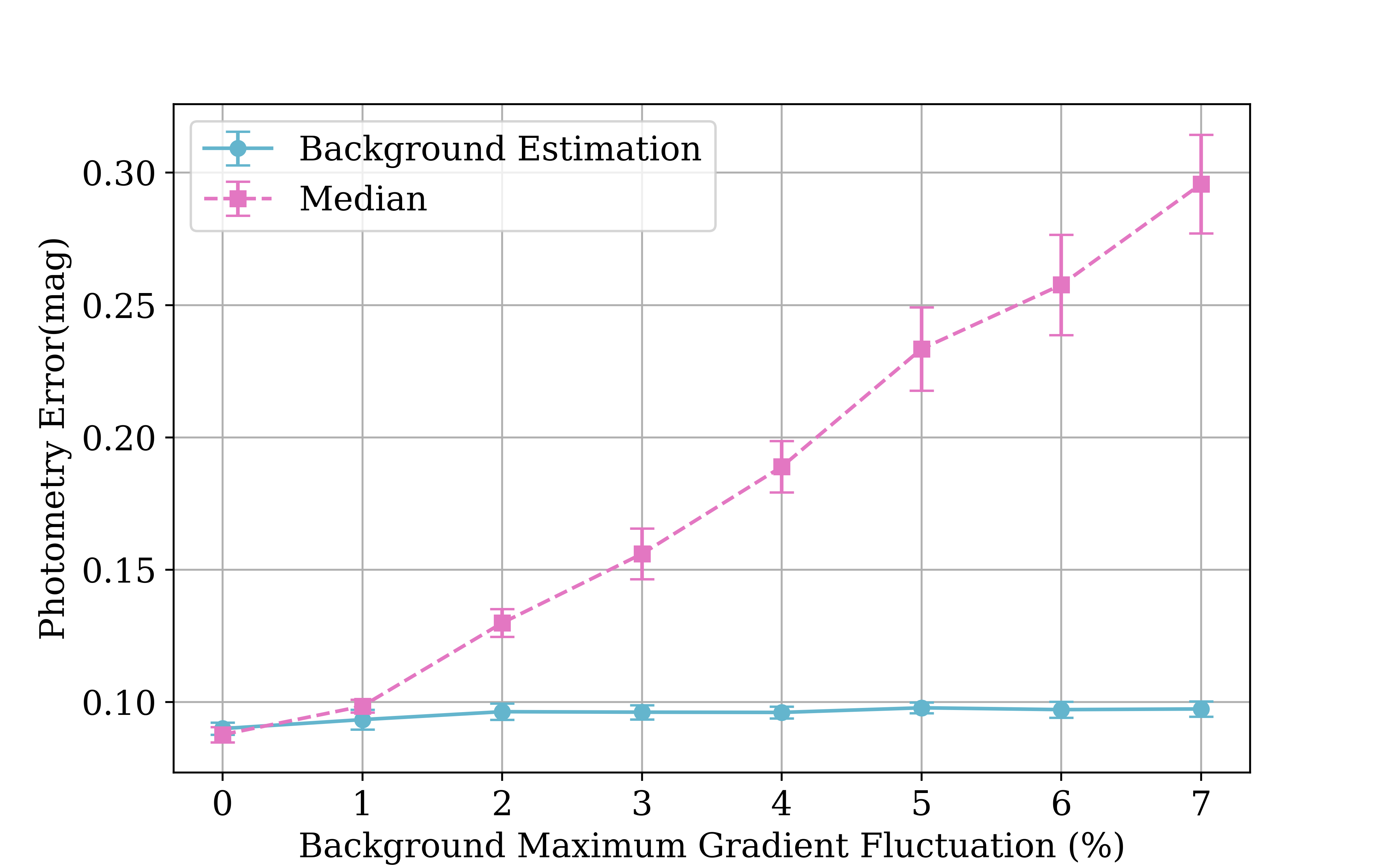} 
        \label{fig:sim_mag_error}
    \end{minipage}
    \caption{The figure demonstrates photometry and detection results obtained from images with different levels of background noises. We can find that simple global median methods cannot adapt to complex background variations, whereas our background estimation function effectively reduces the impact. Panel (a) shows that as the background level increases, the F1-score rate decreases, but with the background estimation and removal methods, we can obtain steady results. Panel (b) shows that the background estimation method effectively reduces photometric measurement errors.}
    \label{fig:sim_back_compare}
\end{figure}

Background noise in astronomical observations arises from multiple sources, including sky brightness, scattered light, and instrumental effects. These disturbances not only affect overall pixel values but also introduce intensity gradients across the image. To assess our processing pipeline's resilience under varying background conditions, we have conducted the following test. Using SkyMaker, we generate multiple star catalogs and designate random stars with magnitudes ranging from 8 to 14 as simulated transient sources. We then create corresponding template and observation images from these catalogs. To replicate realistic background noise patterns, we implement the methodology developed by \citet{jia2015simulation, jia2022digital} to introduce non-uniform background variations, where we have changed the phase interference into gray scale variations. Specifically, we use the mean background level from GWAC observations as the baseline to better reflect actual observational conditions and add a non-uniform background. By adjusting the maximum gradient variation as a percentage of the baseline background level, we systematically control the degree of background noise, allowing us to simulate different levels of background contamination. Following image generation, we apply our image subtraction algorithm to produce difference images, which are then processed using two distinct background removal approaches: our specialized background estimation function and a conventional global median method. The effectiveness of both methods is evaluated using SExtractor as our source detection and photometry tool, allowing for a direct comparison of their respective performances in target detection and measurement.We assess the target detection performance using the F1-score, which is computed from precision and recall as defined previously. For photometric accuracy, we calculate the difference between the measured magnitude and the ground-truth magnitude provided in the input catalog. Figure~\ref{fig:sim_back_compare} presents the results, with the left panel depicting F1-score rate, and the right panel illustrating photometric accuracy.  These results demonstrate that our pipeline maintains high photometric accuracy and robust detection capabilities across a range of simulated observing conditions.

\subsubsection{Performance Evaluation With Different Star Fields}
\label{subsubsec:sex_time_test}
This section evaluates the performance of our source detection algorithm using simulated astronomical images with varying field of views and source densities. Our objective is to assess the algorithm's scalability and efficiency under diverse observational conditions. By systematically altering image dimensions and source counts, we can evaluate its performance in both sparsely and densely populated fields. We generate simulated images using SkyMaker, maintaining consistent background levels, pixel scales, and signal-to-noise ratios while adjusting source density and image size. Source density is controlled by varying the number of sources per square arcminute. For a fair comparison of detection times, we benchmark our custom GPU-based method against SExtractor, excluding the time required for background estimation in the latter to isolate the efficiency of the core source detection algorithms. As demonstrated in Table \ref{tab:gpu_cpu_SEXcomparison}, while both methods exhibit comparable processing performance, the GPU-accelerated approach demonstrates significantly faster detection speeds, particularly when applied to large images with high source densities. This highlights the advantage of GPU acceleration in handling computationally demanding scenarios commonly encountered in astronomical surveys.

\begin{table}
\centering
\caption{Comparison of Source detection time with the SExtractor in CPU and our function in GPU
\label{tab:gpu_cpu_SEXcomparison}}
\begin{tabular}{cccccc}
\hline
\hline
Image Size & Source Density  & GPU Processing Time & CPU Processing Time \\
           &        sources/deg²           & (ms)                & (ms)                \\
\hline
2000x2000 & 50  & 221 & 375 \\
2000x2000 & 100 & 236 & 635 \\
4000x4000 & 50  & 349 & 1541 \\
4000x4000 & 100 & 419 & 2823 \\
8000x8000 & 50  & 826 & 6191 \\
8000x8000 & 100 & 1030 & 10127 \\
\hline
\end{tabular}
\end{table}

\subsection{Performance Evaluation with Real Observation Images}
This section presents a performance evaluation of our pipeline using observational data acquired by the GWAC. We analyze images captured by the telescope's CCD camera with an exposure time of 18 seconds, each covering a field of view of approximately 150 square degrees. Based on the Gaia star catalog, each image contains roughly 20,000 sources with magnitudes less than or equal to 15, corresponding to the detection limit of the GWAC. This results in a high source density predominantly composed of bright celestial objects.\\

We begin by evaluating the performance of our image quality assessment function. The training set of the image quality assement neural network consists of carefully selected images captured during clear, moonless nights, explicitly excluding those affected by artifacts such as shutter shadows. Shutter shadows, caused by imperfections in the camera's mechanical shutter, result in uneven illumination across the image. If not properly addressed, these artifacts can negatively impact subsequent processes, such as image subtraction and source detection, by introducing spurious residuals or false positives. The neural network training process takes 2 hours to complete. After training, we apply the image quality assessment function to real observation images. Processing 200 images, each with dimensions of $4196\times 4136$ pixels, requires just few minutes. Upon manual verification of these images, we identified several low-quality images. Our method has successfully screened out a significant percentage of these low-quality images, with only a small number of false detections. This level of accuracy is deemed acceptable for our applications. Through preliminary testing, our method effectively filters out images with a low Strehl ratio and high background noise. However, since the selection criteria need to be adjusted for different datasets, a direct comparison with traditional methods is not possible, and we have not elaborated further on this part.\\

\begin{figure} 
   \begin{center}
   \begin{tabular}{c} 
   \includegraphics[height=6cm]{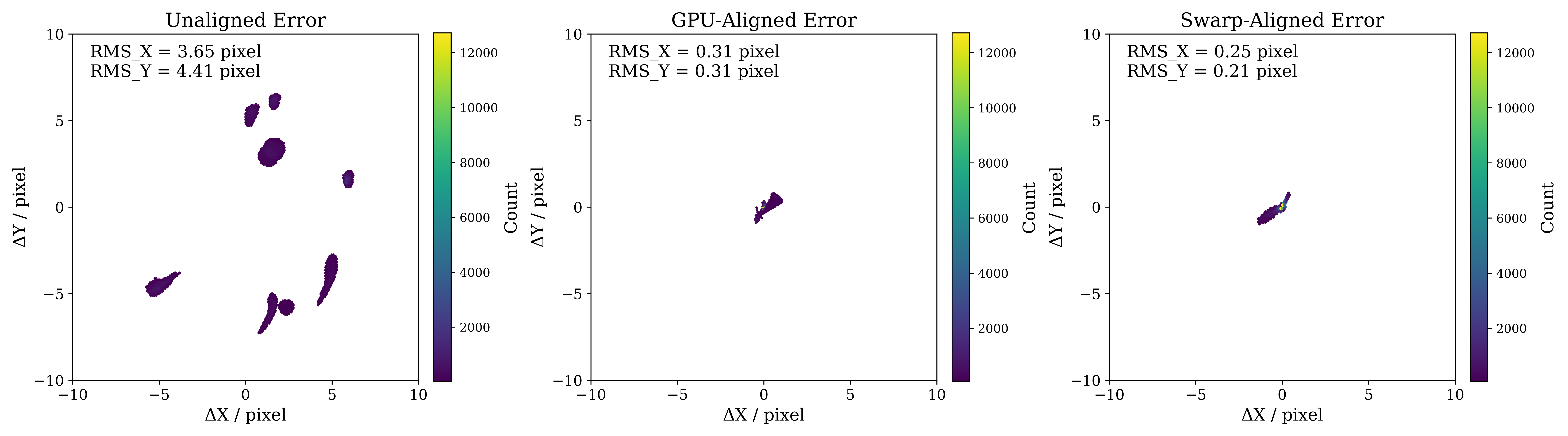}
	\end{tabular}
	\end{center}
   \caption{\label{fig:Imgalign}The residual error comparison between images aligned using Swarp and our GPU based image alignment method shows similar processing results. 
   Our method achieves alignment that is more than 10 times faster while maintaining processing accuracy.}
   \end{figure}

Following the image quality assessment, we proceed to align all images of acceptable quality with those identified as having the best quality in the previous step. Our image alignment function efficiently processes 100 images in just few seconds. For comparison, we also employ Swarp, a traditional image alignment tool, which requires over 130 seconds to process the same 100 images, making our method more than 10 times faster. It's important to note that our pipeline's image alignment function and Swarp utilize different methodologies. To provide a comprehensive evaluation, we present the residual error between stars in aligned images and stars defined by the GAIA DR3 after image alignment for both methods in figure~\ref{fig:Imgalign}. As illustrated, our method achieves faster processing speeds while maintaining similar residual errors compared to the Swarp.\\

\label{sec:align_test}

\begin{figure} 
   \begin{center}
   \begin{tabular}{c} 
   \includegraphics[height=6cm]{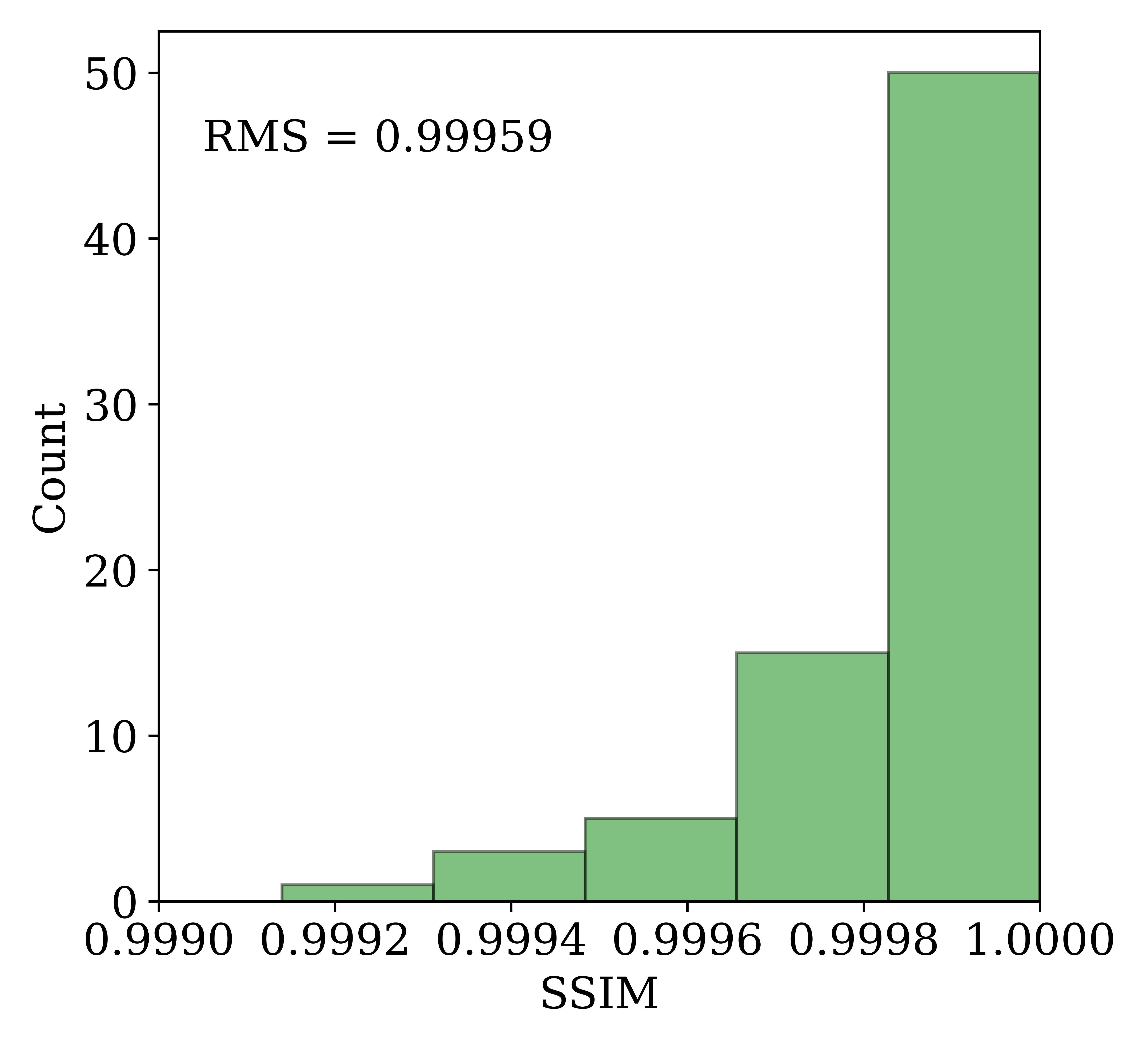}
	\end{tabular}
	\end{center}
   \caption
   { \label{fig:Imgdiff} The histogram of SSIM between difference images obtained by our GPU based image subtraction method and the HOTPANTS. As shown in this figure, our method could obtain similar results of those obtained by HOTPANTS, with times faster.}
   \end{figure}

After aligning all images, we proceed with image subtraction. We employ our GPU based image subtraction function and compare it with HOTPANTS. Our performance evaluation involved processing 100 sets of images. The HOTPANTS algorithm has required several hours to process all images, while our GPU based image subtraction function has completed the task in just a fraction of that time. To assess the quality of our results, we calculated the Structural Similarity Index (SSIM) between the difference images produced by HOTPANTS and our GPU based method, as defined in equation \ref{eq:ssim}:
\begin{equation} \label{eq:ssim}
\text{SSIM}(x, y) = \left( \frac{2\mu_x \mu_y + C_1}{\mu_x^2 + \mu_y^2 + C_1} \right) \cdot \left( \frac{2\sigma_{xy} + C_2}{\sigma_x^2 + \sigma_y^2 + C_2} \right)
\end{equation}
The histogram of these SSIM values is presented in figure~\ref{fig:Imgdiff}. As illustrated, our GPU based image subtraction method demonstrates a significant speed advantage, processing images several times faster than HOTPANTS. Crucially, this substantial increase in speed does not come at the cost of quality. The high SSIM values indicate that our method produces results nearly identical to those of HOTPANTS.\\

Following the image subtraction step, we proceed with background estimation and source detection. For comparison, we employ SExtractor, a widely-used astronomical source extraction software, to estimate the background and perform photometry and astrometry. Concurrently, we utilize the background estimation and source detection functions defined in our framework to execute these same tasks. Processing 100 images with SExtractor requires tens of seconds, while our framework completes the same task in 70\% of that time. This improvement is relatively limited compared to the significant acceleration in other functions. Compared to the significant acceleration observed in other functions, the improvement here is relatively modest. This is because, when processing images with fewer sources, the parallel processing advantage of the GPU is less apparent, and the scale of the computational tasks is too small to fully leverage the GPU's large-scale parallel processing capabilities. For a detailed evaluation of the performance of our GPU-accelerated detection function, please refer to Section \ref{subsubsec:sex_time_test}.\\

\begin{figure}[htbp]
    \centering
    \begin{minipage}[t]{0.43\linewidth}  
        \centering
        \includegraphics[height=1.0\linewidth]{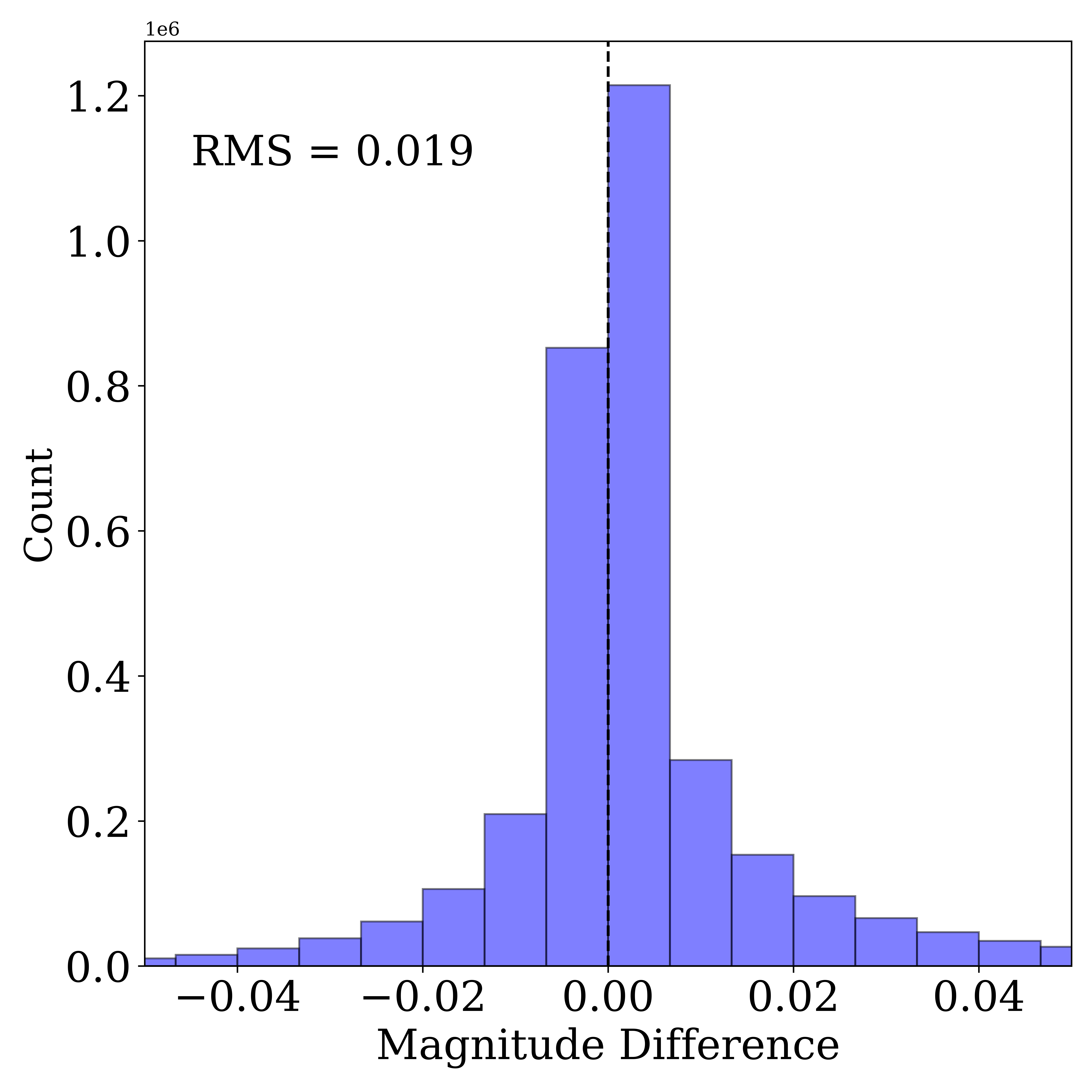}
        \label{fig:mag_diff}
    \end{minipage}
    \begin{minipage}[t]{0.47\linewidth}
        \centering
        \includegraphics[height=1.0\linewidth]{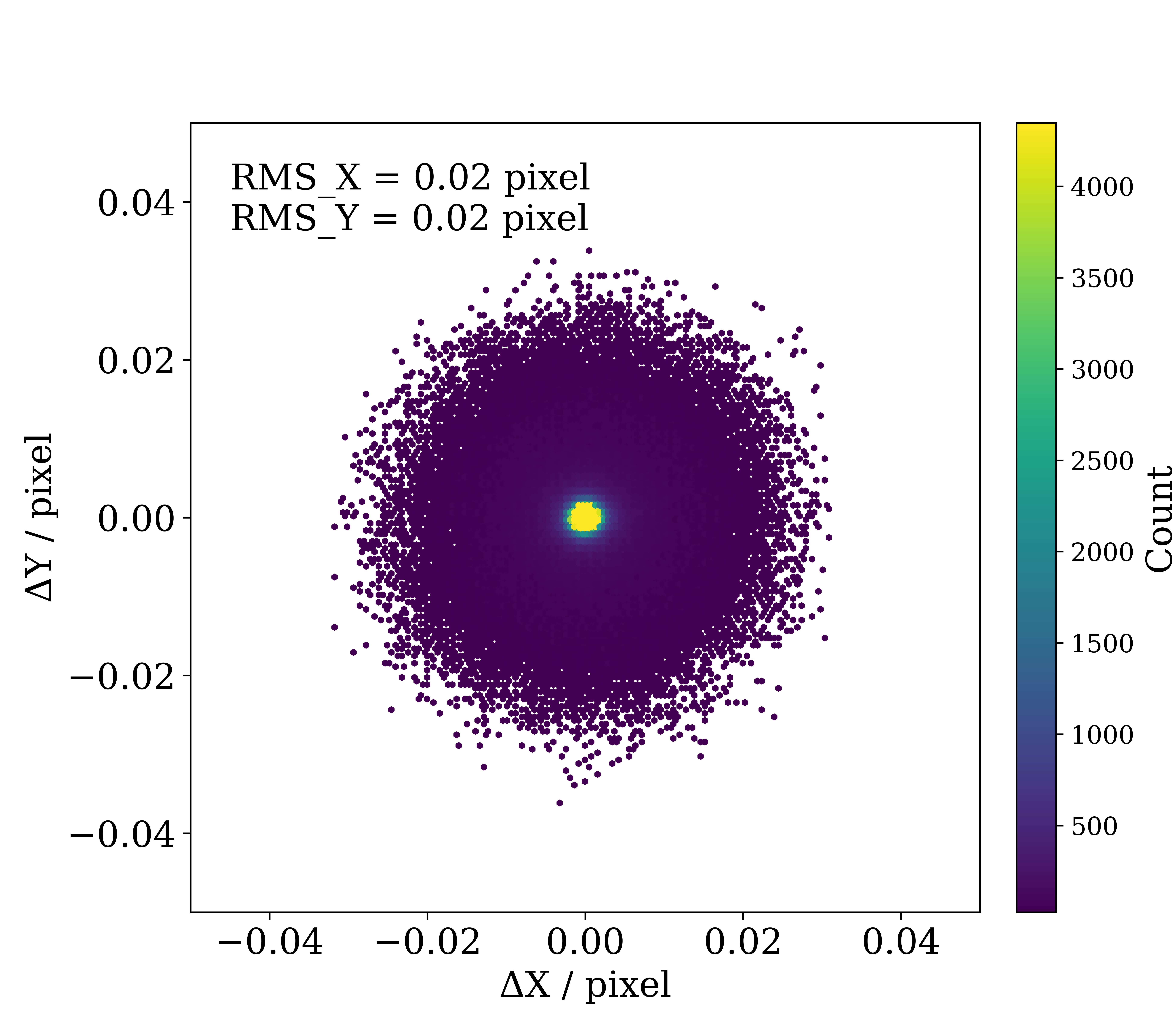} 
        \label{fig:pos_diff}
    \end{minipage}
    \caption{Comparison between our GPU based source detection method and SExtractor in terms of astrometry and photometry information. (a) shows the differences in magnitudes obtained using auto photometry from the two methods, with approximately 95\% of the target photometry differences being less than 0.04. (b) illustrates the differences in source positional information between the two methods, with around 98\% of the target astrometry differences being less than 0.04 pixel.
}
    \label{fig:sex_comparison}
\end{figure}

To evaluate the accuracy of our GPU based function compared to SExtractor, we processed 500 images using identical parameter sets for both methods. The comparative results of astrometry and photometry are presented in figure \ref{fig:sex_comparison}. As evidenced by these figures, our method yields results that are nearly identical to those obtained by SExtractor. The minimal discrepancies observed underscore the high accuracy of our GPU based approach. Notably, while maintaining comparable accuracy, our function demonstrates a significant speed advantage. It processes images approximately 3 times faster than SExtractor. This substantial increase in processing speed, coupled with the preservation of result quality, highlights the efficiency of our GPU-accelerated method in astronomical image analysis.\\

Table~\ref{tab:Processor_compared} presents the average processing time per image for our entire GPU-accelerated image pre-processing pipeline compared to the traditional CPU based approach, as well as the time spent by each function in the process. It is evident that data pre-processing using our framework is more than 12 times faster than traditional methods. The results presented are based on the eager mode of our framework. For the pipeline mode, the main feature is that all parameters are placed into a configuration file for reuse, and, leveraging the NVIDIA DALI framework, it can utilize various GPU-accelerated data augmentation methods from the DALI library to better train the model. However, compared to the eager mode, the pipeline mode did not bring significant improvements in processing performance, and thus we do not discuss it further. Overall, this comprehensive test highlights the versatility and potential of our framework across various astronomical data processing scenarios. By leveraging the advantages of GPU-accelerated functions, our framework offers a powerful solution to modern astronomical data processing challenges, balancing speed, accuracy, and resource optimization.\\

\begin{table}
\centering
\caption{Efficiency of GPU Module Compared to CPU \label{tab:Processor_compared}}
\begin{tabular}{ccccccc}
\hline
\hline
Processor & Image Alignment & Image Subtraction & Background Estimation & Source Extraction & Zscale & Total  \\
       & (ms)            & (ms)             & (ms)                  & (ms)              & (ms)   & (ms) \\
\hline
GPU & 89 & 8187 & 43 & 370 & 18 & 8707 \\
CPU & 1373 & 104152 & 516 & 535 & 141 & 106717\\
\hline
\end{tabular}
\end{table}

\section{Conclusions and Future Works} \label{sec:ConFut}
In recent years, AI based data analysis methods have become increasingly important in time-domain astronomy. These methods require extensive datasets for training or a well-defined pipeline for data preparation. Since most AI methods are developed on GPU architecture, we propose a GPU based image pre-processing framework. This framework includes functions for image quality assessment, background estimation, image alignment, image subtraction, source detection, grayscale transformation, and visualization. To accommodate different scenarios, we have designed our framework with eager mode and pipeline mode. The eager mode is simple and intuitive, while the pipeline mode leverages DALI to achieve fast data loading and intelligent memory allocation. Our tests show that our framework processes data significantly faster than traditional methods while maintaining comparable result quality. This indicates that our framework is highly suitable for time-domain astronomy.\\

To facilitate user adoption, we have packaged the tool into a Docker image, which could be downloaded from the PaperData Repository powered by China-VO. This deployment method not only ensures consistency and stability across various environments but also greatly simplifies the installation and usage process for users. In the future, we will continue to optimize and update the Docker image to support more features and a broader range of application scenarios.\\

\section*{acknowledgement}
We express our gratitude to the reviewer, whose valuable feedbacks have significantly contributed to the improvement of this paper. The code used in this paper will be shared in the PaperData Repository powered by the China-VO. This work is supported by the National Key R \& D Program of China (No. 2023YFF0725300) and the National Natural Science Foundation of China (NSFC) with funding numbers 12173027. This work is supported by the Young Data Scientist Project of the National Astronomical Data Center.\\

\bibliography{sample631}{}
\bibliographystyle{aasjournal}



\end{document}